\documentclass[twocolumn,english,aps,superscriptaddress]{revtex4-2}
\usepackage[T1]{fontenc}
\usepackage[latin9]{inputenc}
\usepackage{amsmath}
\usepackage{graphicx}
\usepackage{esint}

\makeatletter
\usepackage{amsthm}
\usepackage{amsfonts}
\usepackage{bbm}
\usepackage{epsfig}
\usepackage{babel}
\usepackage{color}
\usepackage{framed}
\usepackage{changes}
\usepackage{float}
\usepackage{array}

\usepackage{hyperref}
\hypersetup{colorlinks=true,citecolor=red,linkcolor=blue}

\makeatother

\usepackage{babel}
\begin{document}
\title{Nonlinear self-interaction induced  black hole  bomb}
\author{Cheng-Yong Zhang}
\email{zhangcy@email.jnu.edu.cn}

\affiliation{\textit{College of Physics and Optoelectronic Engineering, Jinan University,
Guangzhou 510632, China }}
\author{Qian Chen}
\email{chenqian192@mails.ucas.ac.cn}

\affiliation{\textit{School of Physical Sciences, University of Chinese Academy
of Sciences, Beijing 100049, China }}
\author{Yuxuan Liu}
\email{223093@csu.edu.cn}

\affiliation{\textit{School of Physics and Electronics, Central South University,
Changsha 418003, China}}
\author{Yu Tian}
\email{ytian@ucas.ac.cn}

\affiliation{\textit{School of Physical Sciences, University of Chinese Academy
of Sciences, Beijing 100049, China }}
\affiliation{\textit{Institute of Theoretical Physics, Chinese Academy of Sciences,
Beijing 100190, China}}

\author{Bin Wang}
\email{wang\_b@sjtu.edu.cn}

\affiliation{\textit{Center for Gravitation and Cosmology, College of Physical
Science and Technology, Yangzhou University, Yangzhou 225009, China}}
\affiliation{\textit{School of Aeronautics and Astronautics, Shanghai Jiao Tong
University, Shanghai 200240, China}}

\author{Hongbao Zhang}
\email{hongbaozhang@bnu.edu.cn}

\affiliation{\textit{Department of Physics, Beijing Normal University, Beijing 100875, China}} 
\affiliation{\textit{Key Laboratory of Multiscale Spin Physics, Ministry of Education, Beijing Normal University, Beijing 100875, China}}
\begin{abstract}
We present the first alternative mechanisms to trigger black hole bomb phenomena beyond the famous superradiant instability. 
By incorporating nonlinear self-interaction into the massive charged scalar field in general relativity, we discover that the allowed static solutions suggest two such novel dynamic mechanisms,  which are further confirmed by our numerical simulations. The first one originates from the linearly unstable hairy black hole, but the bomb can be avoided by dialing the coefficient of the tiny scalar pulse. This distinguishes it from superradiant instability, where the bomb is an inevitable destiny. The second one is an intrinsically nonlinear process, which can even drive a linearly stable Reissner-Nordstr\"om black hole to become a black hole bomb by releasing substantial energy to develop scalar hair. This is also in sharp contrast with superradiant instability which can only drive an unstable black hole. These findings not only open up new avenues for black hole energy burst, but also have potential implications for new phenomena occurring around astrophysical black holes. 
\end{abstract}
\maketitle

{\it Introduction}---Energy extraction from black holes has always been a topic of great interest. Soon after Penrose's proposal that scattering a massive object off a rotating black hole could extract energy \citep{Penrose:1969pc,Penrose:1971uk}, it was shown that a similar process for energy extraction could also be implemented by scattering a charged object off a spherical charged black hole \citep{Denardo:1973pyo}. 
A wave analogue of the Penrose process is superradiance \citep{ZelDovich1971JETP,Misner:1972kx,Press:1972zz,Bekenstein:1973mi}, where the amplified scattering wave carries energy away, causing the black hole to stabilize with slightly less energy \citep{East:2013mfa,Baake:2016oku,Corelli:2021ikv}. 

An intriguing application of superradiance is to make a black hole bomb \citep{Press:1972zz}. With the introduction of a reflecting boundary, the amplified wave can be reflected back towards the black hole. The amplification repeats, generating a superradiantly unstable black hole that continuously releases energy to the wave,  resulting in a sharp growth of the wave. Reflection can be achieved by placing the black hole in an artificial reflecting cavity or in an anti-de Sitter spacetime \citep{Hawking:1999dp}. A more natural approach involves the mass term in the wave equation for a massive bosonic field, which inherently provides an effective reflecting potential barrier \citep{Damour:1976kh}.

With the backreaction of amplified
waves on spacetime, the black hole bomb induced by superradiant instability will inevitably terminate at
some point and the unstable seed black hole must transition to a stable
state. However, exploring the whole progression of this process
is challenging due to the significant disparity in timescales
between the field oscillations and the growth rates of instability \citep{Cardoso:2005vk,Dolan:2007mj,Herdeiro:2013pia,Dolan:2015dha}.
Only a few numerical simulations have successfully
simulated the full evolution of the superradiant instability in asymptotically
flat spacetimes \citep{Sanchis-Gual:2015lje,Sanchis-Gual:2016tcm,East:2017ovw,East:2018glu}
or in anti-de Sitter spacetimes \citep{Bosch:2016vcp,Chesler:2018txn,Chesler:2021ehz}.
The simulations in asymptotically flat spacetimes show that energy can indeed be substantially extracted from the superradiantly unstable black holes, leading to the gradual transition to hairy black holes \citep{Dolan:2015dha,Herdeiro:2016tmi} with smaller energy within the horizon.

For a long time, the creation of black hole bombs was  understood only through the famous superradiant instability. In this {\it Letter}, we present two novel dynamical mechanisms that can also lead to black hole bomb phenomena, even within the simple framework of general relativity minimally coupled with a self-interacting massive charged scalar field.  These mechanisms arise from the nonlinear self-interaction of the matter which is often neglected in studying the superradiant instability at the linear level \citep{Brito:2015oca}.  We first disclose an intriguing linear instability in a branch of hairy black hole  solutions.  Depending on the coefficient of the tiny perturbation, this instability has two possible evolutionary directions. One leads to scalar hair collapse, while the other leads to scalar hair explosion and black hole energy burst.  This makes it  a novel mechanism for triggering a black hole bomb, distinct from the superradiant instability which can only result in hair explosion and energy burst.  
We further demonstrate that Reissner-Nordstr\"om (RN) black holes are always linearly stable in this model, meaning the traditional creation of a black hole bomb from an RN black hole via superradiant instability is absent.  Surprisingly, we can still create a   bomb by injecting a strong scalar pulse into the RN black hole. The nonlinear effect of the scalar field can destroy the stability of the RN black hole and drive it into a hairy black hole by releasing substantial charge and energy to develop scalar hair. This process is intrinsically nonlinear and does not need an artificial reflecting cavity. 
In contrast, superradiant instability can only transform an unstable black hole into a hairy one and requires a reflecting cavity for spherical charged black holes \citep{Sanchis-Gual:2015lje,Sanchis-Gual:2016tcm,Bosch:2016vcp}. 
To our knowledge, these two processes represent the first alternative mechanisms capable of inducing black hole bomb phenomena beyond the  superradiant instability.

 


{\it Model}---The Lagrangian   of the model we consider is (using units where $c=G=1$)
\begin{equation}
\mathcal{L}=R-F_{\mu\nu}F^{\mu\nu}-D^{\mu}\psi(D_{\mu}\psi)^{\ast}-V(\psi),
\end{equation}
where $R$ is the Ricci scalar associated with the metric, the Maxwell field strength $F_{\mu\nu}=\partial_{\mu}A_{\nu}-\partial_{\nu}A_{\mu}$
with $A_{\mu}$ being the gauge potential, and the gauge covariant derivative
$D_{\mu}=\nabla_{\mu}-iqA_{\mu}$ with $q$ the gauge coupling constant
of the complex scalar field $\psi$.  We focus on the potential $V(\psi)=\mu^{2}|\psi|^{2}-\lambda|\psi|^{4}+\nu|\psi|^{6}$
with $\mu$ the scalar field mass, and $\lambda,\nu$ the positive
parameters governing the self-interactions. This potential is widely
used in studying Q-ball \citep{Coleman:1985ki}, a type of non-topological
soliton which may naturally arise in the early universe and is a candidate
for dark matter \citep{Frieman:1988ut,Lee:1991ax,Kusenko:2001vu,Liebling:2012fv}.
It has recently been found that the above nonlinear self-interaction enables this model to circumvent
the no-hair theorem \citep{Mayo:1996mv}, allowing static hairy black
hole solutions \citep{Herdeiro:2020xmb,Hong:2020miv}.

To investigate the static and dynamical properties of the black holes,
we use the spherical Painlev\'e-Gullstrand (PG) coordinates: 
\begin{equation}
ds^{2}=-(1-\zeta^{2})\alpha^{2}dt^{2}+2\alpha\zeta dtdr+dr^{2}+r^{2}d\Omega^{2},
\end{equation}
where $\alpha,\zeta$ are the metric functions dependent on $t$ and $r$.
This coordinate system is regular at the apparent horizon $r_{h}$,
where $\zeta(t,r_{h})=1$. Taking the gauge potential $A_{\mu}dx^{\mu}=Adt$
and introducing auxiliary variables $\Pi=n^{\mu}D_{\mu}\psi$ and
$E=n^{\mu}F_{\mu r}$ with $n^{\mu}=(\alpha^{-1},-\zeta,0,0)$ the unit normal vector to the constant time slice, we get the following constraint equations 
\begin{align}
0= & \partial_{r}E+\frac{2E}{r}-\frac{q}{2}\text{Im}(\Pi\psi^{*}),\label{eq:Bdr}\\
0= & \partial_{r}\zeta+\frac{\zeta}{2r}-\frac{r\left(\rho_{\psi}+E^{2}\right)}{2\zeta}-\frac{r\text{Re}(\Pi^{*}\partial_{r}\psi)}{2},\label{eq:zetadr}\\
0= & \partial_{r}\alpha+\frac{\alpha r\text{Re}(\Pi^{*}\partial_{r}\psi)}{2\zeta},\label{eq:alphadr}\\
0= & \partial_{r}A-\alpha E,\label{eq:Adr}
\end{align}
and evolution equations
\begin{align}
0 & =\partial_{t}\psi-iqA\psi-\alpha(\Pi+\zeta\partial_{r}\psi),\label{eq:psidt}\\
0 & =\partial_{t}\Pi-\frac{\partial_{r}\left(\alpha\left(\Pi\zeta+\partial_{r}\psi\right)r^{2}\right)}{r^{2}}-iA\Pi q+\alpha\frac{\partial V}{\partial\psi^{*}}.\label{eq:pidt}
\end{align}
Here $\rho_{\psi}=T_{\mu\nu}^{\psi}n^{\mu}n^{\nu}=(|\Pi|^{2}+|\partial_{r}\psi|^{2}+V)/2$
is the projection of the scalar field stress-energy tensor along $n^\mu$. Given the initial distribution of $\psi$ and $\Pi$, we obtain $E,\zeta,\alpha$ and $A$ by solving the constraint equations (\ref{eq:Bdr},\ref{eq:zetadr},\ref{eq:alphadr},\ref{eq:Adr})
successively. The evolution equations (\ref{eq:psidt},\ref{eq:pidt})
are then solved by Runge-Kutta method to get $\Pi,\psi$ on the next
time slice. By repeating this procedure, we can obtain the matter
and metric data on all time slices. 

At spatial infinity, Minkowski spacetime should be approached. This implies the following boundary conditions for solving the equations:
\begin{equation}
\zeta\to\sqrt{\frac{2M}{r}},\ A\to\Phi+\frac{Q}{r},\ E\to-\frac{Q}{r^{2}},\ \psi,\Pi\to0,\label{eq:BC}
\end{equation}
where $M$ and $Q$ are the total mass and charge of the system, respectively.
$\Phi$ is the gauge potential and we take $\Phi=0$ hereafter.
We also set $\alpha=1$ at infinity by fixing the auxiliary freedom
of $\alpha dt$ in PG coordinates.

We   trace
the dynamical evolution  by the scalar field energy  $E_{\psi}=\frac{1}{4\pi}\int_{r_{h}}^{\infty}dV\rho_{\psi}$, the black hole charge $Q_{h}=\frac{1}{4\pi}\oint_{r_{h}}dSF_{\mu\nu}n^{\mu}s^{\nu}$
with $s^{\nu}$ the outward pointing unit normal vector to the apparent horizon two-sphere \citep{Torres:2014fga}, as well as the black hole mass, defined by the Christodoulou-Ruffini formula as 
$M_{B}=M_{h}+\frac{Q_{h}^{2}}{4M_{h}}$ with $M_{h}=\sqrt{S_{h}/16\pi}$ the irreducible mass and $S_{h}=4\pi r_{h}^{2}$ the apparent horizon area \citep{Christodoulou:1971pcn,East:2013mfa,Baake:2016oku,East:2017ovw,East:2018glu,Corelli:2021ikv}. There should be charge conservation $Q_h+Q_\psi=Q$ during dynamical evolution, where
$Q_{\psi}=\frac{1}{4\pi}\int dVn^{\mu}j_{\mu}$ with
$j_{\mu}=-\frac{q}{2}\text{Im}(\psi^{*}D_{\mu}\psi)$ the charge current carried by the scalar field. This fact can be employed to monitor our numerical simulations.




{\it Bombs suggested by static solutions}---As alluded to before, due to the nonlinear self-interaction, the model we are considering allows not only the  RN  solution with $\psi=0$, but also hairy
black hole solutions. The hairy black hole solutions have a static geometry and stress-energy
tensor while the scalar field oscillates as $\psi(t,r)=\phi(r)e^{i\omega t+if(r)}$
in PG coordinates. Here $f(r)\equiv q\chi(r)-\omega g(r)$ is a phase. It can be removed by a combination of the gauge transformation $\psi\to\psi e^{-iq\chi},A\to A-\partial_{r}\chi$
and the coordinate transformation $t\to t+g(r)$. The scalar field exhibits an
asymptotic behavior $\phi\to\frac{c_0e^{-\mu_{\infty}r}}{r}$ at spatial infinity,
where $c_0$ is an irrelevant constant and $\mu_{\infty}\equiv\sqrt{\mu^{2}-\omega^{2}}$
is the effective mass of the complex scalar field. The bound $\omega\le\mu$
should be satisfied for static hairy solutions; otherwise the
scalar field would have infinite total energy and thus unphysical. Regularity on the horizon requires that the frequency
$\omega$ satisfies the resonance condition $\omega=qA(r_{h})$.
Given 
the gauge coupling $q$, black
hole horizon radius $r_{h}$ and total charge $Q$, the static hairy
solutions can be worked out by solving the static equations of motion with Newton-Raphson
method. Other quantities such as $\omega,Q_{h}$ and $M_{B}$ can
be derived from these solutions.


\begin{figure}[h]
\begin{centering}
\includegraphics[width=0.99\linewidth]{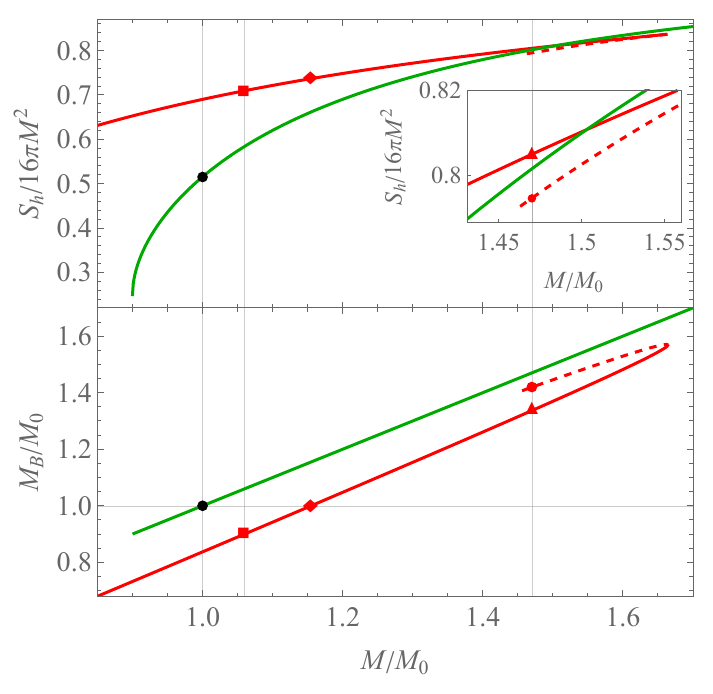}
\par\end{centering}
{\footnotesize{}\caption{{\footnotesize{}\label{fig:static} The black hole area $S_h$ and mass $M_B$ versus the total mass of the static
 black hole solutions with fixed  $Q=0.9M_0$ when $qM_0=3$. The green line denotes the RN solutions, while the solid and dashed red lines represent the stable and unstable hairy black holes, respectively.
The red and black points indicate the seed black holes in Figure \ref{fig:UnstableHBH} and \ref{fig:energyExtraction},
respectively.}}
}{\footnotesize\par}
\end{figure}

Hereafter, we assume a reference RN black hole with a total mass $M_{0}=1$ and a total charge $Q_{0}=0.9M_{0}$. The parameter  $M_{0}$ serves to establish the energy scale of the problem. Then we select the black hole solutions with a total charge $Q$ identical to that of the reference black hole, but the total mass $M$ can vary. The reason we fix $Q=Q_{0}$ is to facilitate comparison with results from dynamical simulations later, where the total charge remains unchanged while the total mass $M$ increases with perturbation strength.   Without loss of generality, we further fix $qM_0=3$ and the  potential $V(\psi)=\frac{|\psi|^{2}}{M_0^2}(1-\frac{|\psi|^{2}}{0.1^{2}})^{2}$ which satisfies the weak and dominant energy conditions. Then the results for selected solutions with $Q=Q_0$ are shown in Figure \ref{fig:static}. Alongside the RN solution, two additional branches of hairy solutions emerge, coinciding at a certain maximum value of $M$.
As displayed later, the dashed branch turns out to be linearly unstable, while the solid red branch and the RN branch  are linearly stable. 
By organizing static solutions in such a fresh manner,  we can readily argue for the feasibility of making black hole bombs through two potentially distinct mechanisms from the familiar superradiant instability.



Let us first start from the red point with $M=1.47M_0,M_B=1.42M_0$ on the dashed branch. Since this seed hairy black hole is linearly unstable, even a slight perturbation can drive it to one of two possible stable final states at $M=1.47M_0$: either the stable hairy black hole on solid red branch or the RN black hole on the green line. This speculation is supported by the fact that both of them have a larger horizon area $S_h$ than the seed black hole. If the stable hairy black hole, as indicated by the red triangle, is the final state, then the unstable seed black hole must release substantial energy outward to transition to it, since the final state has a smaller black hole mass $M_B$ within the horizon compared to the seed one. This implies a black hole bomb made from a linearly unstable hairy black hole.  


On the other hand, it is also possible to make a black hole bomb out of an RN solution,  say the black point with $M=M_B=M_0$ on the green line. Although this seed RN black hole is linearly stable under a small perturbation, its stability under a large perturbation is not guaranteed. In particular, if we engineer the scalar field away from the realm of the validity of linear perturbation theory, with its charge density vanishing outside of the horizon, then according to the Penrose inequality ($M\ge M_B$) \citep{Hayward:1998jj,Disconzi:2012es,McCormick:2019fie}, we can inject energy to make the total mass $M$ of the system lie between $M_0$ and $1.155M_0$, the value determined by the intersection diamond point of the horizontal gray line and the solid red curve. If this process can trigger the transition from the black point to a hairy black hole on the solid branch with $M_0<M<1.155M_0$,  represented by the red square, then the linearly stable seed RN black hole with $M_B=M_0$ must release a substantial amount of energy outward, since the final hairy black hole has $M_B<M_0$ within the horizon. By this way, we can create a black hole bomb even from a linearly stable seed black hole.
This intrinsically nonlinear process is permitted by the non-decreasing area law since the red square has a larger horizon area than the black point.

Below we shall confirm the above two novel dynamical mechanisms to make black hole bombs by our fully nonlinear numerical simulations, whereby we also find that both of them are controllable.




\begin{figure}[h]
\begin{centering}
\includegraphics[width=0.96\linewidth]{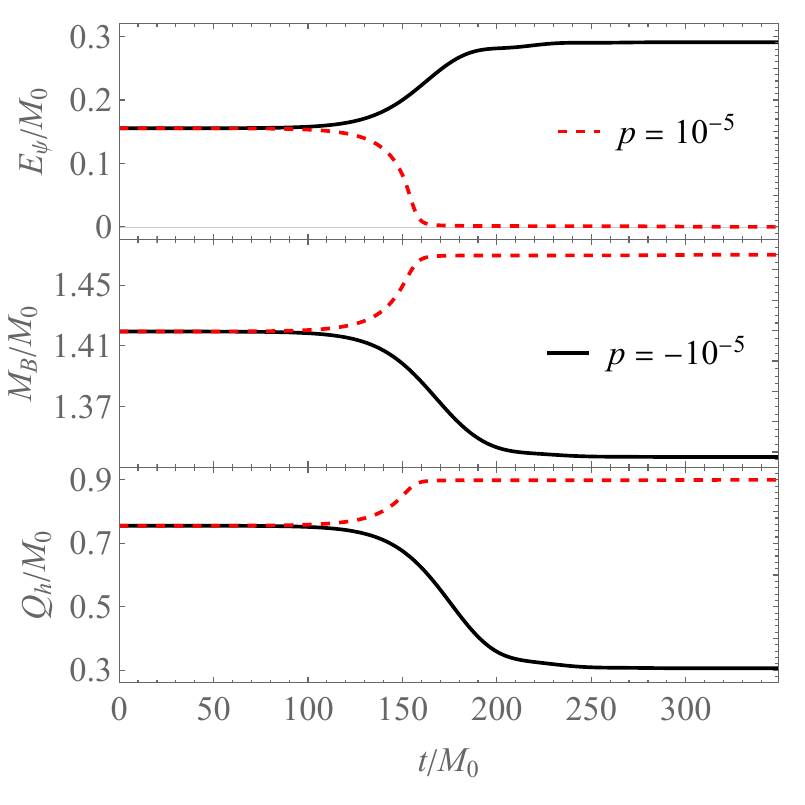}
\par\end{centering}
{\footnotesize{}\caption{{\footnotesize{}\label{fig:UnstableHBH}The evolution of scalar field energy $E_{\psi}$, black hole
mass $M_{B}$ and charge $Q_{h}$ 
after the perturbation (\ref{eq:pert0}) with $p=\pm 10^{-5}$ of the unstable hairy black hole with $M=1.47M_0,M_B=1.42M_0$ (red point in Figure \ref{fig:static}). The corresponding total mass increment after perturbation
 is about $10^{-7}M$.}}
}{\footnotesize\par}
\end{figure}
{\it Bombs confirmed by dynamical evolution}---We first confirm that the black hole bomb can be made out of the unstable hairy black hole. As such, we perturb the red point in Figure \ref{fig:static} with the following ingoing pulse
\begin{equation}
    \delta\psi=pe^{-\frac{M_0}{r-r_{1}}-\frac{M_0}{r_{2}-r}}\frac{(r-r_{1})(r_{2}-r)}{M_0^2},\ \delta\Pi=\partial_{r}\delta\psi,
    \label{eq:pert0}
    \end{equation}
if $r_{1}<r<r_{2}$ and zero otherwise. We fix $r_{1}=4M_0,r_{2}=9M_0$ in the simulations. With a tiny $p$, the increase in total mass is negligible.  
Typical
simulations are shown in Figure \ref{fig:UnstableHBH}. Depending
on whether $p$ is positive or negative, the unstable seed hairy black hole
evolves into either an RN black hole or a linearly stable hairy black hole. When it evolves into the latter, the seed black hole releases a substantial amount of charge and energy into the scalar  hair, similar to the black hole bomb triggered by superradiant instability. 
However, superradiant instability always results in scalar hair explosion and energy extraction from the black hole. Unlike this, here when the unstable seed hairy black hole instead evolves into an RN black hole, the scalar hair collapses rather than explodes, and its energy is absorbed by the black hole. This makes this linear instability a novel mechanism for triggering a black hole bomb beyond the superradiant instability.
It is worth pointing out that this novel mechanism has the potential to extract up to $25\%$ of the black hole energy, which occurs with a linearly unstable hairy extremal black hole with $Q=M=0.9M_0$ for smaller $q$ such as $qM_0=1$.
  


\begin{figure}[h]
\begin{centering}
\includegraphics[width=0.96\linewidth]{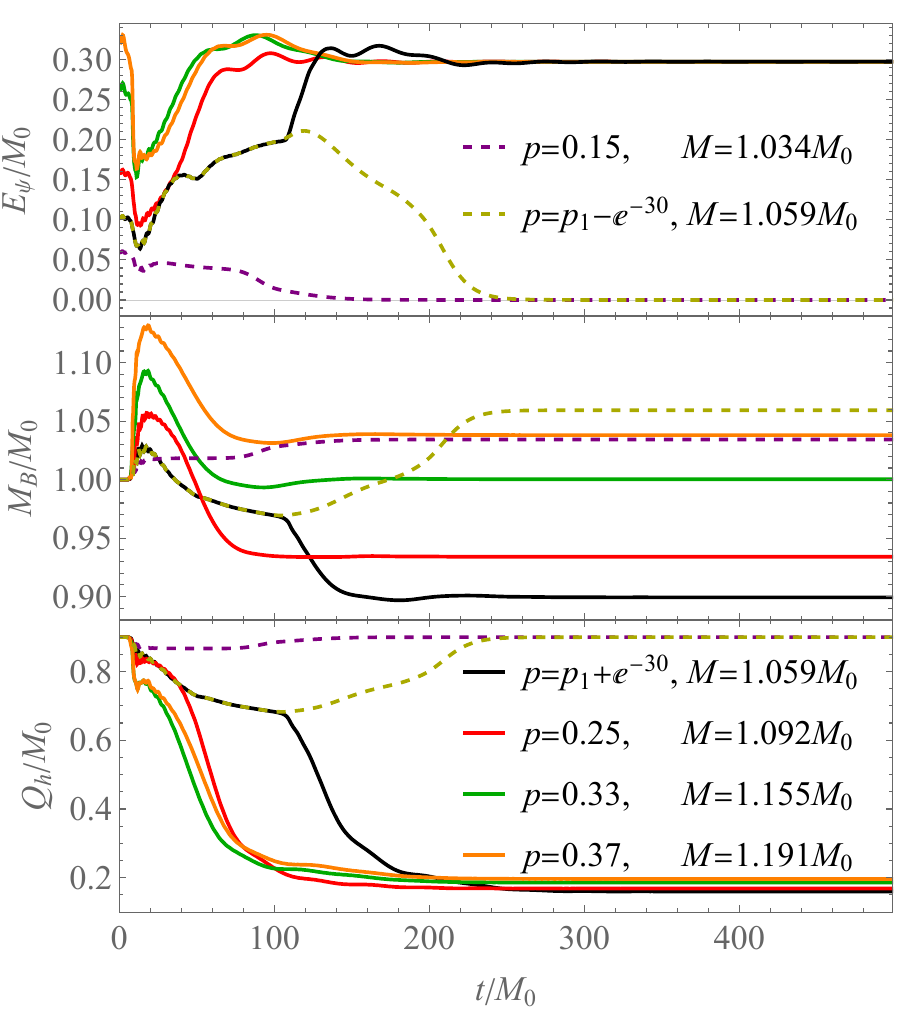}
\par\end{centering}
{\footnotesize{}\caption{{\footnotesize{}\label{fig:energyExtraction} The evolution of scalar field energy $E_{\psi}$, black
hole mass $M_{B}$ and charge $Q_{h}$ from an RN black hole with $M=M_B=M_0$ (black point in Figure \ref{fig:static}) under pulse injection (\ref{eq:pert}).
The dashed curves represent the evolution resulting in RN black holes,
while the solid curves represent the evolution resulting in hairy
black holes. The threshold $p_{1}\approx0.198609115264384$ is
determined with an accuracy up to order $10^{-15}$ using the bisection
method.}}
}{\footnotesize\par}
\end{figure}
To show that the bomb can also be made out of a linearly stable RN black hole, we inject the black point in Figure \ref{fig:static} with the following ingoing  scalar pulse:
\begin{equation}
\delta\psi=0.1pe^{-(\frac{r-12M_0}{2M_0})^{2}},\ \delta\Pi=\partial_{r}\delta\psi.\label{eq:pert}
\end{equation}
The total mass increases with amplitude $p$, while the total charge is unchanged since the initial pulse has  vanishing current $\delta j_{\mu}$ everywhere.
We plot the relevant result in
Figure \ref{fig:energyExtraction}. For pulse (\ref{eq:pert}), 
there exists a threshold $p_{1}$. 
When $p<p_{1}$, the scalar pulse is finally absorbed by the seed RN black hole, leading to another RN black hole with a larger black hole mass. 
This result is in accord with the fact that the RN black hole under consideration is linearly stable. Thus one cannot create a  bomb from an RN black hole through the traditional superradiant instability in this model. However, for $p>p_{1}$, the strong nonlinear effect of the scalar field destroys the linear stability of the RN black hole and drives it to transition into a hairy black hole.
In particular, when $p_{1}<p<0.33$,  the final hairy black hole has a smaller mass $M_{B}$ within the horizon than the seed black hole. Actually, up to $10\%$ of the black hole energy and $80\%$ of the black hole charge were extracted in our numerical simulations. This implies that we can  create a black hole bomb in the nonlinear regime from an RN black hole, even it is linearly stable. 
But when $p>0.33$, the total mass exceeds $1.155M_0$, so as illustrated in Figure \ref{fig:energyExtraction} for $p=0.37$, the final hairy black hole has a larger mass $M_B$ than the seed black hole, which means no black hole bomb for larger $p$. 

Therefore, we have fulfilled our promise made before. In particular, we find that the bomb made via either of the two novel mechanisms is controllable in the sense that it can be triggered only within the partial region of the parameter space of the scalar  pulse profile.

{\it Conclusion}---Due to the nonlinear self-interaction of the massive charged scalar field, our model allows for not only RN solutions, but also two branches of hairy black hole solutions. This together with the non-decreasing area law as well as Penrose inequality suggests two novel dynamical mechanisms to create a black hole bomb, which is further confirmed explicitly by our fully nonlinear numerical simulations. The first one originates from the linearly unstable hairy black hole, but to be a bomb requires the coefficient of the scalar pulse profile be negative, which makes it distinct from the superradiant instability, whereby the bomb is unavoidable at all. The second one also differs from the superradiant instability, since it is triggered intrinsically in the nonlinear regime, where the strong nonlinear self-interaction of scalar field can even drive a linearly stable RN black hole to transition into a hairy black hole by releasing substantial energy to develop scalar hair. As far as we know, these two processes are the first alternative mechanisms to the superradiant instability for creating a black hole bomb. 



These findings highlight the significant impact of nonlinear self-interaction of matter fields on black hole energy burst.  There are indications that analogous phenomena widely exist for black holes with other  self-interacting massive solitary hairs,
such as Proca and axion hair \citep{SalazarLandea:2016bys}, which will be studied elsewhere in the future. On the other hand, given the mass gap between hairy and bald rotating black hole solutions \citep{Herdeiro:2014pka,Herdeiro:2015tia,Brihaye:2014nba}, a tiny perturbation cannot transform a Kerr black hole to the hairy one. But the second mechanism which is intrinsically nonlinear as we disclosed here may apply to these black holes. In particular, with the involved matter field as a candidate for dark matter in mind,  we have reason to expect that the vast energy burst induced by the novel mechanisms could produce intriguing observational signatures, which await to be further explored.


\section*{Acknowledgments}

This work is supported by the Natural Science Foundation of China (NSFC)
under Grant No. 12035016, 12075026, 12075202, 12375048, 12375058, and 12361141825, as well as the National Key R\&D Program of China under Grant No. 2021YFC2203001.
\bibliographystyle{apsrev4-2ideal}
\bibliography{BHQhair}


\section*{Supplemental Material}

We include supplemental material to demonstrate the results
with varying parameters and scalar field potentials. This material
unveils additional intriguing phenomena observed in the dynamical
simulations and emphasizes the generality of our findings.

\begin{itemize}
\item \textit{Simulations with a larger $q$}
\end{itemize}

In the main text, we worked with $V(\psi)=\frac{|\psi|^{2}}{M_{0}^{2}}(1-\frac{|\psi|^{2}}{0.1^{2}})^{2}$
and $qM_{0}=3$. The results from these settings have led us to propose
novel mechanisms for black hole bomb phenomena. Here we simulate the
evolution with a larger $q$ and showcase other intriguing dynamical
phenomena that emerge during the evolution. 

\begin{figure}[H]
\begin{centering}
\includegraphics[width=1\linewidth]{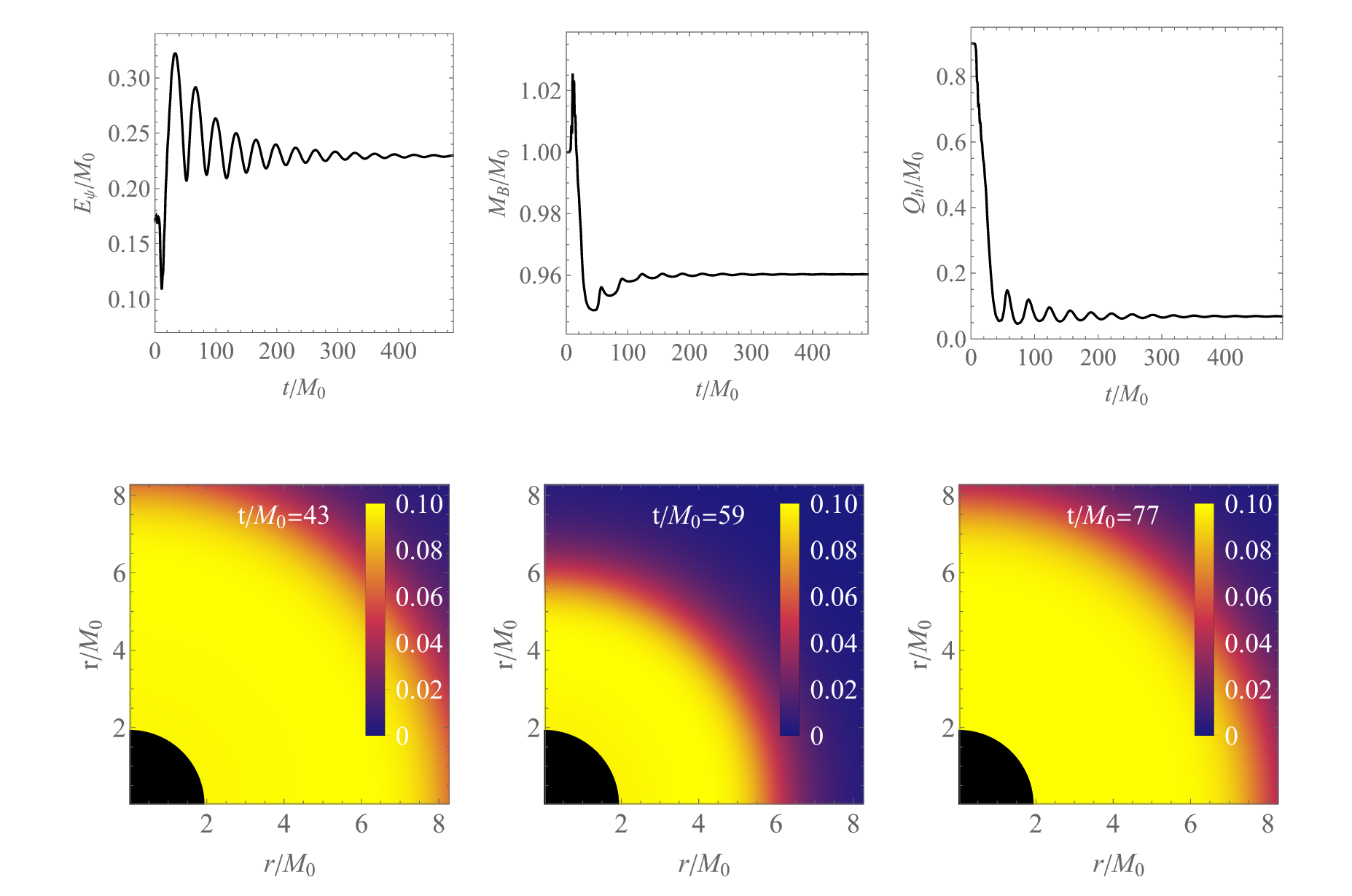}
\par\end{centering}
{\footnotesize{}\caption{{\footnotesize{}\label{fig:largeqOsi} }\textbf{\footnotesize{}Upper
panel}{\footnotesize{}: The evolution of the scalar field energy $E_{\psi}$,
the black hole mass $M_{B}$ and charge $Q_{h}$ starting from an
RN black hole with $M_{0}=1,Q=0.9M_{0}$ under the perturbation $\delta\psi=0.1pe^{-(\frac{r-12M_{0}}{2M_{0}})^{2}},\delta\Pi=\partial_{r}\delta\psi$
with $p=0.26$ when $qM_{0}=5$. }\textbf{\footnotesize{}Lower panel}{\footnotesize{}:
Snapshots of the absolute value $|\psi|$ of the scalar field during
the evolution. The black regions in the lower left corner represent
the region inside the apparent horizon at the corresponding times.}}
}{\footnotesize\par}
\end{figure}

In Figure \ref{fig:largeqOsi}, we present the dynamical simulations using the same
potential but with the parameter $qM_{0}=5$. In the upper panel,
we observe that the scalar field energy $E_{\psi}$, black hole mass
$M_{B}$ and charge $Q_{h}$ all experience rhythmic growths and declines
over a long period. As depicted in the lower panel, this behavior
is accompanied by the scalar field displaying prolonged rhythmic radial
expansions and contractions during the evolution, reminiscent of the
radial pulsations of Cepheid variable stars. To our knowledge, this
is the first observation of such behavior in a black hole system.
The radial expansions and contractions of the scalar field are due
to the intense competition between the gravitational attraction and
self-interaction versus the repulsion from energy extraction and electric
forces. It suggests a time-dependent geometry that has not yet stabilized.
In each cycle, a small amount of scalar field energy is absorbed by
the central black hole. Eventually, a hairy black hole forms, characterized
by a static geometry and stress-energy tensor while the scalar field
continues to oscillate. Details of the final solutions are elaborated
in the section \textquotedbl\textit{Bombs suggested by static solutions}\textquotedbl{}
in the main text.

\begin{itemize}
\item \textit{Simulations with other potential }
\end{itemize}

We have focused on the potentials of the form $V(\psi)=\mu^{2}|\psi|^{2}(1-\frac{|\psi|^{2}}{\sigma^{2}})^{2}$. However,
the novel black hole bomb mechanisms we have discovered are not confined
to this specific potential.

\begin{figure}[H]
\begin{centering}
\includegraphics[width=1\linewidth]{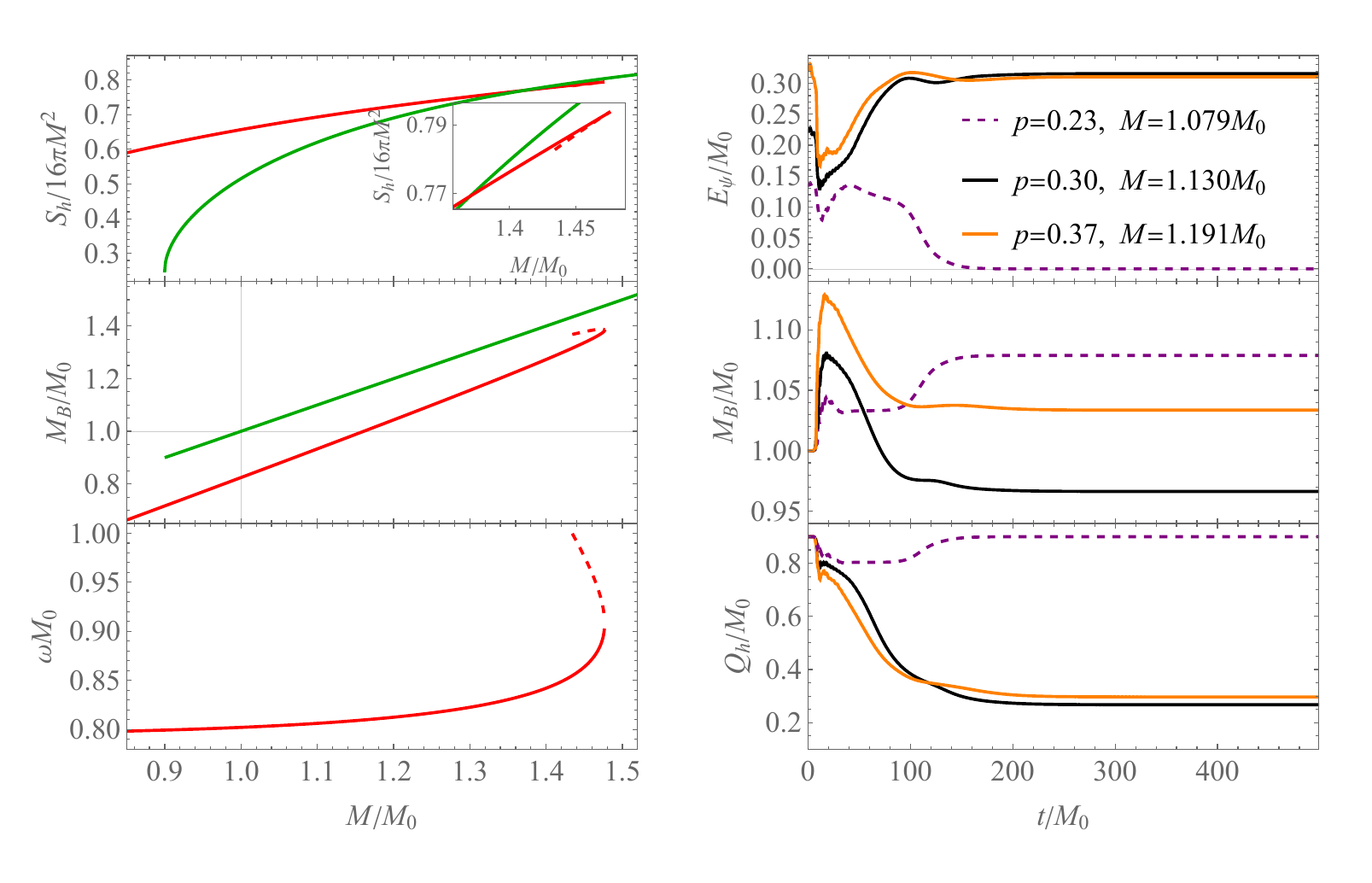}
\par\end{centering}
{\footnotesize{}\caption{{\footnotesize{}\label{fig:potentialEvo} }\textbf{\footnotesize{}Left
panel}{\footnotesize{}: Static black hole solutions with $V(\psi)=\frac{|\psi|^{2}}{M_{0}^{2}}(1-\frac{2|\psi|^{2}}{0.1^{2}}+\frac{1.1|\psi|^{4}}{0.1^{4}})$
and $qM_{0}=3$. The green line denotes the RN solutions, while the
solid and dashed red lines represent the stable and unstable hairy
black holes, respectively. The unstable hairy branch terminate at
$\omega M_{0}=1$ since the bound $\omega\le\mu=M_{0}^{-1}$ should
be satisfied. All static solutions in the left panel have the same
charge with a reference RN black hole which has a total mass $M=M_{0}$
and a total charge $Q=0.9M_{0}$. }\textbf{\footnotesize{}Right panel}{\footnotesize{}:
The evolution of the scalar field energy $E_{\psi}$, the black hole
mass $M_{B}$ and charge $Q_{h}$ starting from the reference RN black
hole under the perturbation $\delta\psi=0.1pe^{-(\frac{r-12M_{0}}{2M_{0}})^{2}},\delta\Pi=\partial_{r}\delta\psi$
when $qM_{0}=3$.}}
}{\footnotesize\par}
\end{figure}

In Figure \ref{fig:potentialEvo}, we present results using a different
potential $V(\psi)=\frac{|\psi|^{2}}{M_{0}^{2}}(1-\frac{2|\psi|^{2}}{0.1^{2}}+\frac{1.1|\psi|^{4}}{0.1^{4}})$
with $qM_{0}=3$. The left panel shows static black hole solutions
with three branches, similar to those in Figure 1 of the main text,
indicating that the dynamical features are qualitatively similar across
different scalar potentials. In the right panel, we indeed observe
the same qualitative behavior as in Figure 3 of the main text: a sufficiently
strong perturbation (with $p=0.3$) can trigger a black hole bomb
from the linearly stable RN black hole, but an overly large perturbation
(with $p=0.37$) results in no net energy extraction.

Additionally, selecting a seed black hole from the linearly unstable
hairy black holes on the dashed branch leads to similar dynamical
behaviors as shown in Figure 2 of the main text. For simplicity, we
do not include these results here.

We have also tested the potential $V(\psi)=\frac{|\psi|^{2}}{M_{0}^{2}}(1-\frac{1.9|\psi|^{2}}{0.1^{2}}+\frac{|\psi|^{4}}{0.1^{4}})$
and obtained similar results. Note that all tested potentials satisfy
the energy conditions. These experiments reinforce the generality
of our findings across various potentials. 

\end{document}